\documentclass[aps,prl,showpacs,reprint,superscriptaddress,superscriptaddress,twocolumn]{revtex4-1}
\usepackage{graphicx}  
\usepackage{dcolumn}   
\usepackage{bm}        
\usepackage{amssymb}   
\usepackage{amsmath}
\usepackage{epstopdf}
\usepackage{color}
\usepackage{appendix}
\begin{document}

\widetext

\title{Robust finite-temperature disordered Mott insulating phases in inhomogeneous Fermi-Fermi mixtures with density and mass imbalance}

\author{Anzi Hu}
\affiliation{Joint Quantum Institute, University of Maryland and National Institute of Standard and Technology, Gaithersburg, Maryland 20899, USA}
\author{M. M. Ma\'ska}
\affiliation{Department of Theoretical Physics, Institute of Physics, University of Silesia, PL-40007 Katowice, Poland}
\author{Charles W. Clark}
\affiliation{Joint Quantum Institute, University of Maryland and National Institute of Standard and Technology, Gaithersburg, Maryland 20899, USA}
\author{J. K. Freericks}
\affiliation{Department of Physics, Georgetown University, Washington, D.C. 20057, USA }

\date{\today}
\begin{abstract}
Ultracold mixtures of different atomic species have great promise for realizing novel many-body phenomena. In a binary mixture of femions with a large mass difference and repulsive interspecies interactions, a disordered Mott insulator phase can occur. This phase displays an incompressible total density, although the relative density remain compressible. We use strong-coupling and Monte Carlo calculations to show that this phase exists for a broad parameter region for ultracold gases confined in a harmonic trap on a three-dimensional optical lattice, for experimentally accessible values of the trap parameters. 
\end{abstract}
\maketitle


The advancement of ultracold atom experiments with fermionic mixtures \cite{jordens2008mott, Kohstall2012,liao2010spin,greif2013short,trotzky2008time,giorgini2008theory,jo2009itinerant}, particularly the recent realization of a metastable fermionic mixture of $^{40}$K-$^6$Li, demonstrates the possibility of realizing a quantum degenerate binary fermonic mixture with great mass and density imbalance. Previous theoretical studies have focused on ground-state or low-temperature phases of the repulsive imbalanced fermion mixtures, discussing the segregated phase or itinerant ferromagnetism in density imbalanced systems \cite{Duine2005,MaskaFreericks2008, Simons2009, jo2009itinerant, conduit2009itinerant, Troyer2010, Paramekanti2009, chang2011ferromagnetism, Lun2013} and the crystallization and complex long range density ordering in mass imbalanced systems \cite{brandt1991phase,Freericks2002,Freericks2002-2,Freericks1996,Freericks1999,Freericks2003,MaskaFreericks2008}. These complex phases are often unstable to thermal fluctuations and the inhomogeneity of the trapping potential and they have yet to be realized in experiments.

It is of great importance to explore new many-body phenomena that are robust against thermal fluctuations and inhomogeneity. Here we show an example of such a robust many-body phase: a disordered
Mott insulator (DMI) phase in a mixture of localized and itinerant fermions. This phase \emph{only} exists at finite temperatures that are sufficiently large for thermal fluctuations to suppress density segregation \cite{DongenLeinung1997, Freericks2003}. At low temperatures, a trapped mixture tends to phase separate and the DMI phase no longer exists \cite{MaskaFreericks2008}.
We show that in an inhomogeneous system, the DMI phase exhibits an incompressible total density while the relative density is compressible. The DMI also exists in cases of large number imbalance and asymmetries in the trapping potentials for the two fermonic species. 
It can be detected with procedures similar to those used for detecting a Mott insulator phase of fermionic $^{40}$K in Refs. \cite{jordens2008mott, schneider2008metallic} or the dual Mott insulator phase in a Bose-Fermi mixture of Yb isotopes \cite{sugawa2011interaction}. 

We treat a mixture of $N_f$ heavy and $N_c$ light fermions in a cubic lattice with lattice constant $a$ and additional isotropic harmonic potentials, at finite temperature $T>0$. 
Due to the mass asymmetry, $M_f>M_c$, there is a large difference in the hopping energies of the two species. Thus, the system can be effectively described by a model of localized and itinerant fermions \cite{Ates_Ziegler2005,MaskaFreericks2008}. 
Its Hamiltonian is written as

\begin{eqnarray}
H =&-&J\sum_{\langle j,j'\rangle} c^{\dagger}_{j}c_{j'}+U\sum_{j} c^{\dagger}_j c_j f^{\dagger}_j f_j \nonumber \\
  &+&\sum_{j}\left[(V_{c,j} -\mu_c)c^{\dagger}_j c_j + (V_{f,j} -\mu_f)f^{\dagger}_j f_j \right],%
\label{hamiltonian}
\end{eqnarray}
where $c_j$ and $f_j$ are the annihilation operators for light and heavy fermions at lattice site $j$, $J$ is the hopping energy for the light fermions, $U>0$ is the repulsive interaction between light and heavy fermions, $\mu_{c/f}$ is the chemical potential for light/heavy fermions and $V_{c/f,j}$ is the corresponding harmonic trapping potential at lattice site $j$, specifically $V_{c/f,j}=(M_{c/f}/2)\omega^2_{c/f}r^2_j$ where $\omega_{c/f}$ is the trapping frequency for heavy/light fermions and $r_j$ is the distance of the lattice site $j$ from the center of the trap. It is convenient to define $L_{c/f}$ through $V_{c/f,j}=J(r^2_j/L_{c/f})^2$. The total particle number of each species, $N_{c/f}$, is determined by the chemical potentials $\mu_{c/f}$. We use both the strong coupling (SC) method \cite{Hu2011} and Monte Carlo (MC) calculations to investigate the finite-temperature phases of these mixtures. The supplementary material \cite{SMMC} contains details of the SC derivations and the MC calculations and demonstrates their agreement for the parameter region considered in this letter.

\begin{figure}[h!]
\includegraphics[width=0.5\textwidth]{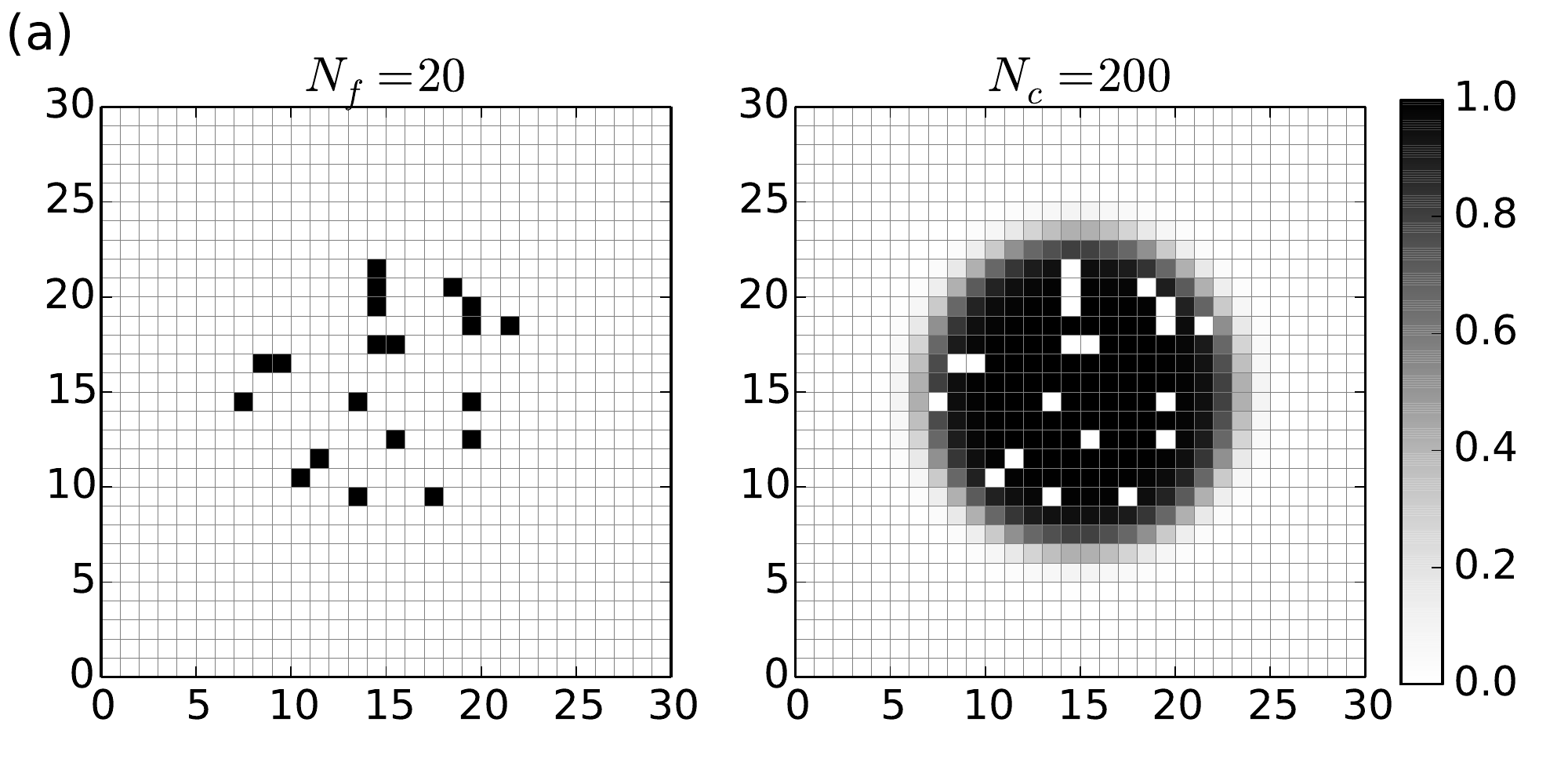}
\includegraphics[width=0.5\textwidth]{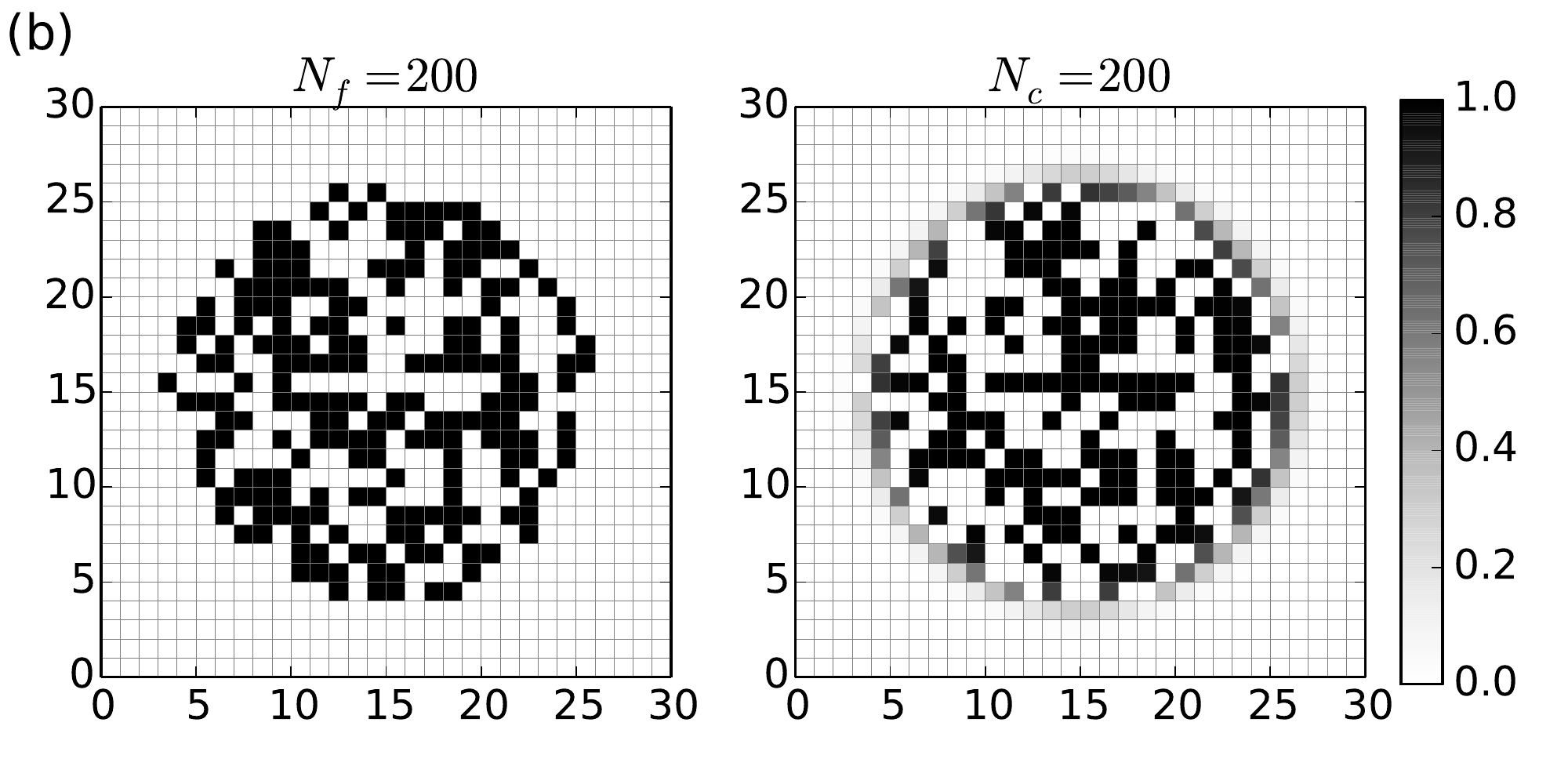}
\includegraphics[width=0.5\textwidth]{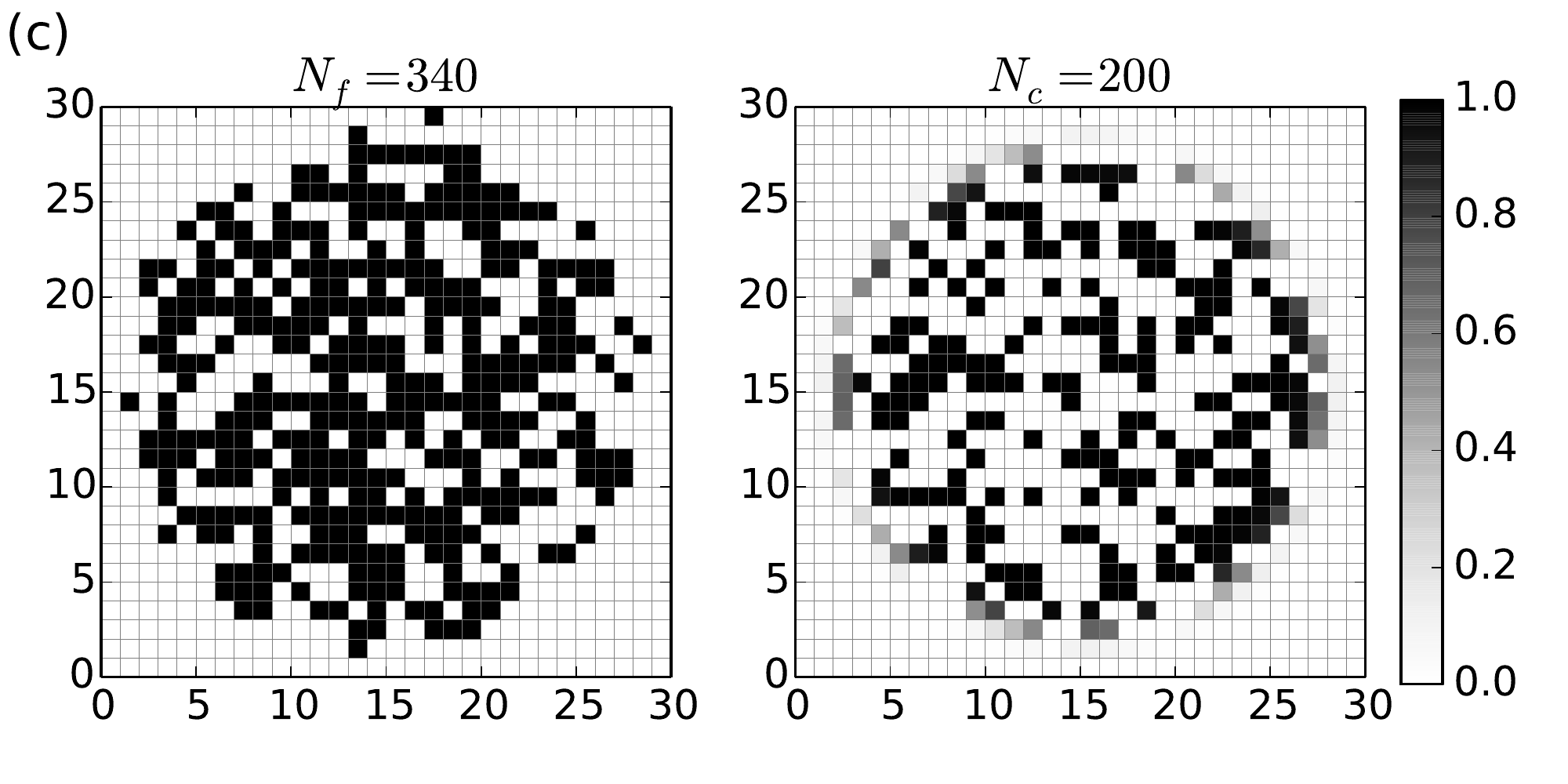}
\caption{Monte Carlo snapshots of the disordered MI phase in a trapped two-dimensional lattice system of light and heavy mixtures. The light particle number is fixed at 200 and the heavy particle numbers are (a) 20, (b) 200, and (c) 340. The interaction $U$ is $50J$, the trap frequencies are set by $L_c=L_f=2a$ and the temperature is $2J/k_B$. The density distribution of the light and heavy particle appears disordered but the sum of them, the total density, remains at unit filling near the center for all the cases \cite{SMMC}. This plateau of the total density in the center of the trap remains in the MI phase.}
\label{fig:F0}
\end{figure}

An intrinsic feature of  ultra-cold atom experiments is the spatial inhomogeneity induced by the existence of a trapping potential. With the MC calculation, we simulate experimental in-situ images of atoms in trapped systems through snapshots generated by the MC simulation (after thermalization) \cite{MaskaFreericks2008}. These snapshots show a striking feature of the DMI. Figure \ref{fig:F0} shows series of possible experimental realizations of the DMI phases for three particular choices of $N_f/N_c$. The snapshots of the density distribution are generated by a MC simulation for a two-dimensional lattice in a harmonic trap. We consider the process of adding heavy fermions as impurities into an ultra-cold gas of light fermions, with a strong repulsive interaction between the light and heavy fermions. If no heavy fermions are present, the light fermions form a band insulator at the center of the trap as a result of the trapping potential. When heavy fermions are added, one heavy fermion leads to the modification of the total wave function of the light fermions. In the three panels shown, the individual density distribution is disordered. But when we compare the distributions of the two atomic species, they are perfectly complimentary. Around the center of the trap, the total density \emph{always} remains at unit filling \cite{SMMC}. The radius of the MI plateau in the total density increases as more heavy particles are added, but the MI plateau remains for a very large range of particle number ratios. 
This re-configuration of both fermions is the result of the strong anti-correlation due to their interactions. If there was no interaction, the light fermions would still form a band insulator when heavy fermions are added. The added heavy fermion will form a separate Fermi gas with either a compressible or incompressible state depending on the trapping potential and particle number. Since there is a tendency at low temperature for the two species to phase separate in a homogeneous mixture \cite{freericks2002segregation}, the inhomogeneity of the trap and the thermal energy at a finite temperature both act to stabilize this MI phase.




\begin{figure}[h]
\includegraphics[width=0.45\textwidth]{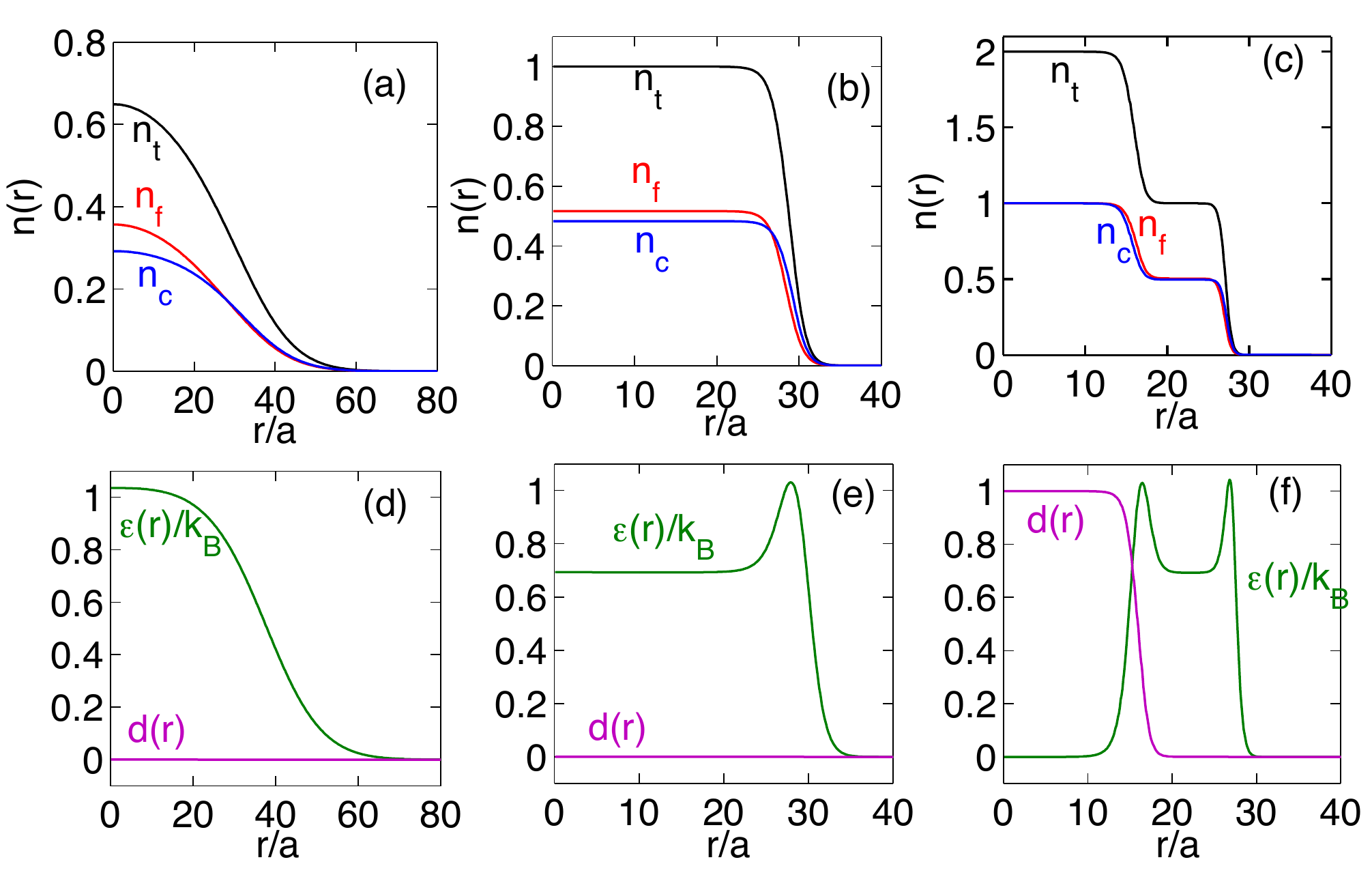}
\caption{(Color online) Radial distribution of densities, entropy and double occupancy for different trapping frequencies in a 3D system. The interaction is fixed at $U=50J$ and temperature at $T=2J$. There are $5\times10^4$ heavy and light fermions. The trapping frequencies are the same for both heavy and light fermions, $L_c=L_f=L$. (a)-(c): Radial distribution of the total density $n_t$ (magenta line), heavy fermions $n_f$ (red line) and light fermions $n_c$ (blue line). (d)-(f): Radial distribution of the entropy $\epsilon (r)$ (green line) and the double occupancy $d(r)$ (black line). (a) and (d): The trapping frequency is set by $L=16.5a$. The system is in the metallic state and the density is compressible. (b) and (e): The trapping frequency is set by $L=5.0a$. A MI state is developed at the center. A peak is formed in the entropy distribution $\epsilon(r)$ at the edge of the MI phase, because the strong anti-correlation of the MI leads to a reduction of local entropy. (c) and (f): The trapping frequency is set by $L=3.0a$. A band insulator is formed at the center and a ring of the MI exists near $r=20a$. A metallic state exists elsewhere.} 
\label{fig:F1}
\end{figure}

\begin{figure}[h]
\includegraphics[width=0.45\textwidth]{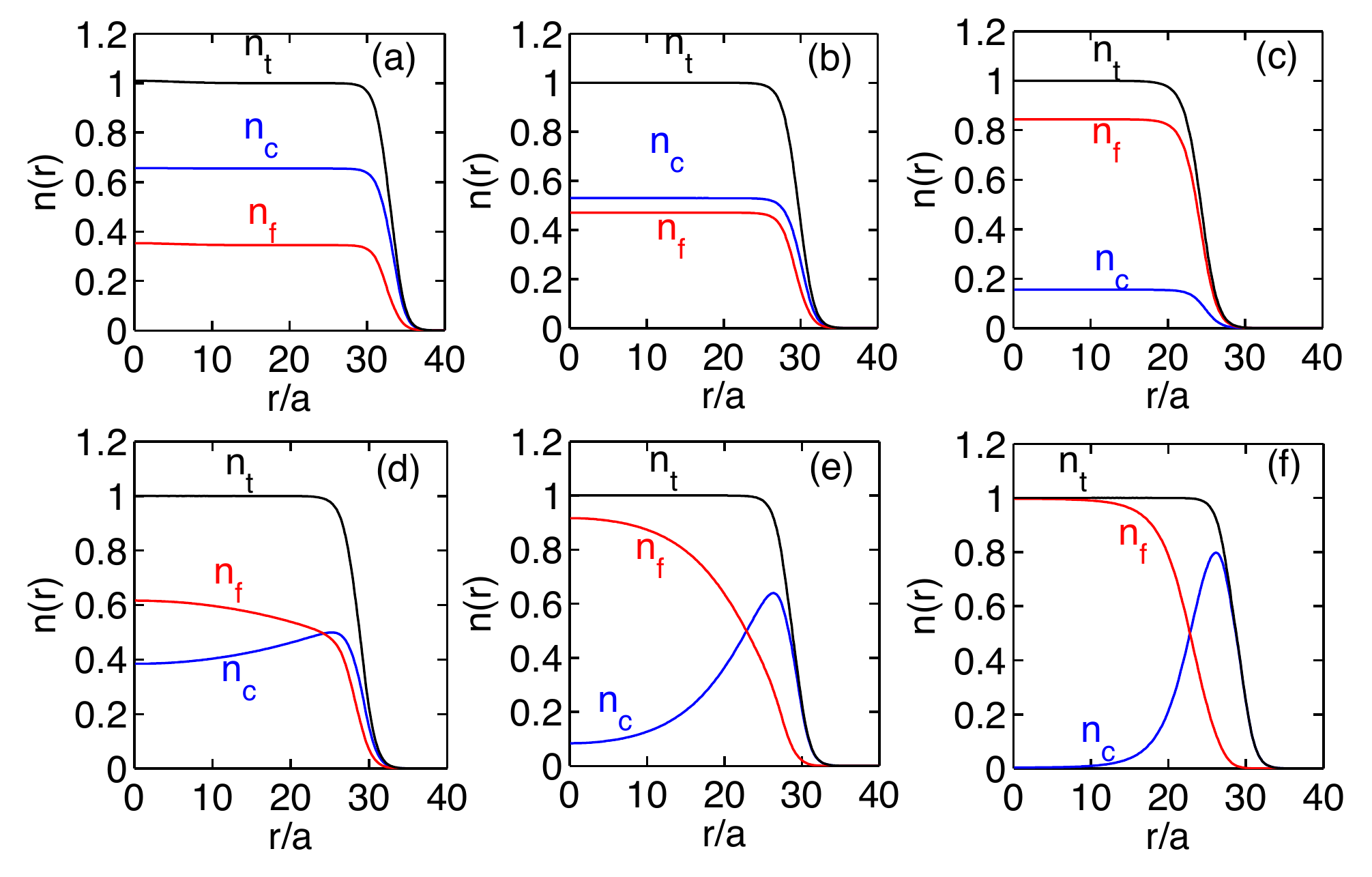}
\caption{(Color Online)Radial density distribution for systems of asymmetric particle numbers and trap potentials. The interaction is fixed at $U=50J$ and temperature at $T=2J$. (a)-(c): changing the light particle number while keeping the heavy particle number fixed at $5\times10^4$ and trap potentials for both fixed at $L_c=L_f=L=5a$. The light particle number is $1\times10^5$ (a), $6\times10^4$ (b) and $1\times10^4$ (c). Remarkably, the system remains in a MI phase for all the range of the particle numbers with the total density fixed at unity while the individual densities form a plateau at various fractional fillings. (d)-(f): changing the trap potential of the heavy fermions while keeping the particle number fixed at $5\times10^4$ for both and the trap potential for the light fermions fixed at $L_c=5a$. The trapping potential for the heavy fermions is set by $L_f=4.9a$ (d), $L_f=4.5a$ (e) and $L_f=4a$ (f). }
\label{fig:F4}
\end{figure}

\begin{figure}[h]
\includegraphics[width=0.5\textwidth]{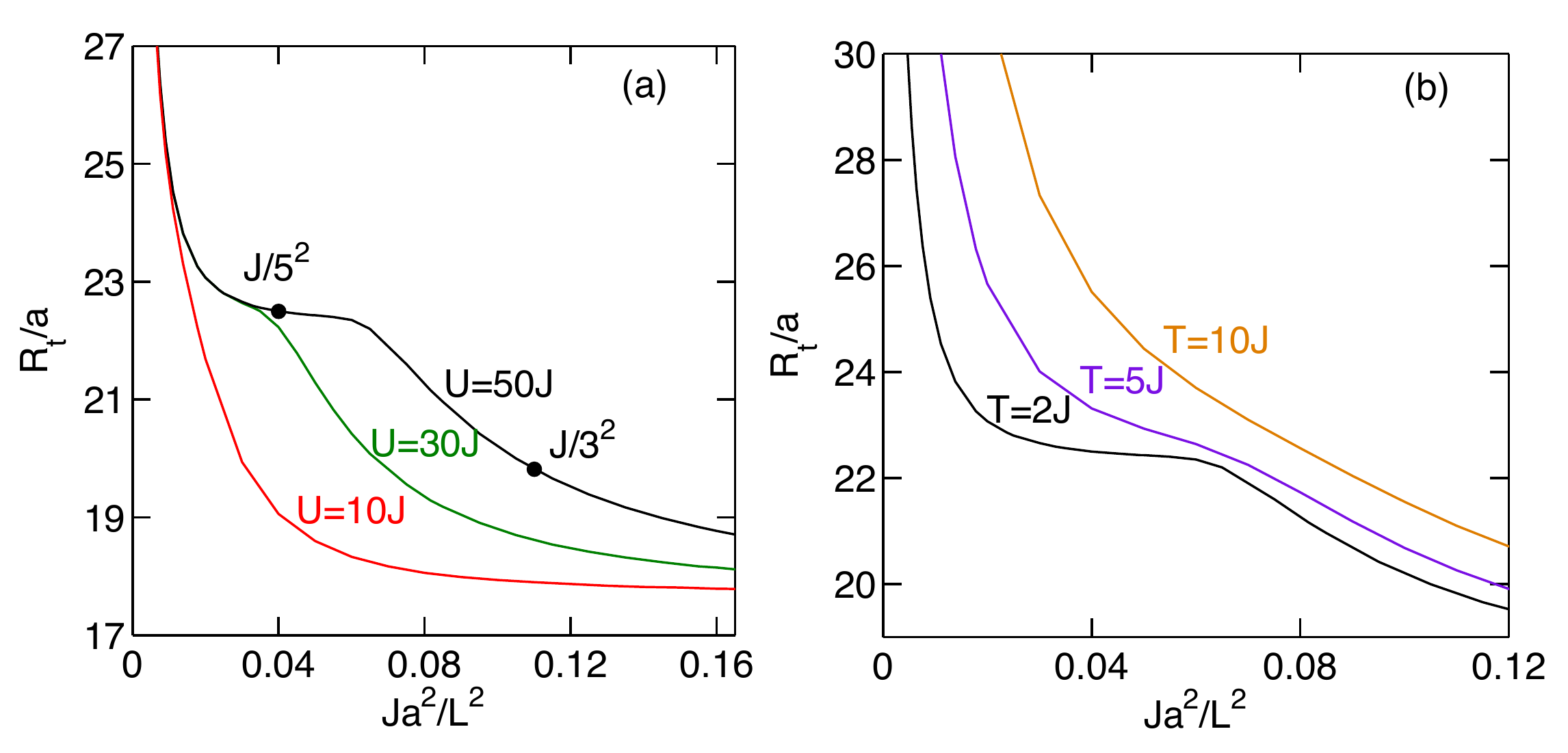}
\caption{(Color online) Cloud size $R_t$ as a function of the trapping frequency for different interaction $U$ (a) and different temperatures $T$ (b). The particle number of each species is fixed at $5\times10^4$. (a): Temperature is fixed at $2J$ and $U$ changes from $10J$ to $50J$.  The black dots corresponds to the MI and band insulator cases in Fig. \ref{fig:F1}. (b): Interaction $U$ is fixed at $50J$ and temperature varies from $2J$ to $10J$.}
\label{fig:F2}
\end{figure}

With the SC method, we calculate much larger systems for three-dimensional inhomogeneous systems. We find that the inhomogeneity leads to complex spatial dependence of different phases. In Fig.~\ref{fig:F1}, we show that the radial distributions of the densities, entropy and double occupancy for a strongly interacting system undergo different phases as the trapping curvature changes. Here, the double occupancy, $d_j=\langle c^{\dagger}_jc_jf^{\dagger}_jf_j\rangle$, expresses the probability of having both the light and heavy atom on site $j$,. 
In Figs.~\ref{fig:F1} (a) and (d), we show the case of a shallow potential 
($L=16.5a$). The cloud expands to minimize the kinetic energy and the total density is less than one. The local entropy changes with the density and the double occupancy is almost zero due to the strong repulsive interaction.  As the trapping potential becomes strong, the cloud is forced inwards and the density at the center of the trap increases. When the total density reaches unit filling, the Hubbard band gap prevents it from increasing its density further and a plateau of unit filling is formed. In this region, the cloud size stays largely unchanged as the trap curvature increases. In Figs.~\ref{fig:F1} (b) and (e), we show the case of an incompressible MI phase at the center of the trap. It is important to note that in this MI phase, the incompressibility only applies to the total density and the individual densities can still be compressible, which is further demonstrated in Fig. \ref{fig:F4}. The unit filling of the total density is \emph{not} the result of each species forming an incompressible phase at half-filling, but the strong anti-correlation between the light and heavy particles that guarantees there is always \emph{either a light or heavy particle}. This anti-correlation leads to a reduction of the local entropy, which is most noticeable at the edge of the MI plateau. At the edge, there is a MI and metallic state  for almost identical densities. However, the entropy in the metallic state is much higher because the light and heavy particles are less correlated. This leads to a peak in the entropy distribution about the edge. In Figs. \ref{fig:F1} (c) and (f), we discuss the case where the trapping potential is strong enough to force the particle to fill in the upper Hubbard band and a band insulator is formed for both species at the center of trap. Away from the center, there is a secondary plateau that corresponds to the MI phase. In Fig. \ref{fig:F1} (f), we show distinct behavior of the entropy and double occupancy for different phases. From the center of the trap, the band insulator state is characterized by a sharp increase of the double occupancy to one and a sharp decrease of the entropy to zero. The metallic state is characterized by an increase of the local entropy. The MI state is characterized by a plateau of the local entropy which is reduced from the entropy of the surrounding metallic state. The double occupancy in both the metallic and the MI state is extremely low as a result of the strong interaction.

The anti-correlation between the particles in the MI phase is further illustrated under asymmetric conditions. First, we consider the case of large particle number asymmetry. As is shown in Fig. \ref{fig:F0} of the MC calculation of the 2D lattice, the MI phase is robust for large number asymmetry. The robustness is confirmed in the SC calculation for 3D systems.  We change the light particle number from $1\times10^4$ to $1\times10^5$. Remarkably, the system \emph{always} self-organizes into the MI phase, even under extreme particle number asymmetry. In Figs. \ref{fig:F4} (a), (b) and (c), we show the radial density distribution for the cases of $N_c=1\times10^5$, $6\times10^4$, and $1\times10^4$. In all three cases, a MI plateau is present at the center of the trap. The plateau is signaled by unit filling of the total density. The radius of the plateau is different in each case because the total particle number is different. For $N_c=1\times10^5$, the individual density forms a plateau at $n_c\approx0.65$ and $n_f\approx0.35$. For $N_c=6\times10^5$, the individual density forms a plateau at $n_c\approx0.55$ and $n_f\approx0.45$. For $N_c=1\times10^4$, the individual density forms a plateau at $n_c\approx0.15$ and $n_f\approx0.85$. The same behavior is observed when we fix the light particle number while changing the heavy particle number. The remarkable robustness of such MI phases also points to a new possibility of creating a density plateau at any fractional filling of a species by changing the other species' particle number. We next consider the case of trap frequency asymmetry by varying the trap potential for the heavy particles while keeping the light one fixed via the choice $L_f=5a$. In the case of asymmetric traps, the difference of the trap potential causes particles to reorganize to minimize their energy. Because the relative density remains compressible, it changes responding to the difference in the local chemical potential. 
This is the case for Fig. \ref{fig:F4}(d) and (e). When the difference of the trap potentials is too large, the particles are spatially separated. This is the case in Fig. \ref{fig:F4} (f), where the center of trap becomes a band insulator of only heavy particles and the light particles are forced outside the band insulator. Note that this phase separation is different from the phase separation at much lower temperature, which happens for symmetric traps and particle numbers \cite{MaskaFreericks2008}. This phase separation is induced by the asymmetry of the trap potential. 

In experiment, the MI phase can be detected from both the mixture's cloud size and the double occupancy \cite{SMMC}  following procedures similar to previous experiments with single species of atoms\cite{jordens2008mott, schneider2008metallic}. 
The cloud size of the mixture is determined based on the total radial density profile as $R_t=\sqrt{(N_c+N_f)^{-1}\sum_j r^2_j n_t(r_j)}$. Figure \ref{fig:F2} shows the cloud size as a function of the trapping frequency for different interaction strengths and different temperatures for a system of $5\times10^4$ light and heavy particles.  In Fig. \ref{fig:F2} (a), we consider the dependence of the MI phase on the interaction strengths at a given temperature $T=2J$.  When the interaction is relatively small ($U=10J$), the cloud size decreases as the trap potential increases until a band insulator is formed at the center. When the interaction is large ($U=30J, 50J$), there exists a plateau in the cloud size that corresponds to the MI phase formed at the center of the trap. The plateau appears around $U=30J$ and grows for stronger interactions. In Fig. \ref{fig:F2}(b), we consider the MI phase at different temperatures with $U=50J$. We find that the critical temperature is around $T=5J$. Above $T=5J$, the cloud size decreases smoothly with an increase of the trap potential. When $T<5J$, the decrease is slowed down corresponds to the MI phase. For $T=2J$, a plateau is clearly present. 

It is worth noting that the critical interaction and temperature indicated in our calculation of a trapped system is not quantitatively the same with those calculated for a homogeneous system. This is because in the trapped system, the Hubbard gap needs to be compared with the potential energy gradient of the trap. If the gap is so small that the energy difference between neighboring sites at the center of the trap and the thermal fluctuations are sufficient to overcome the gap, a MI plateau will not form. Hence, the measurement of macroscopic quantities, such as the cloud size and average double occupancy rate, is more appropriate to detect the existence of a collective MI region in the trapped system, instead of a measuring of the critical interaction strength for the corresponding homogeneous system.


In summary, we have demonstrated that despite the fragility of the magnetic ordering in mixtures of atoms with large mass differences, the Mott phase
exists at a relative high temperature and in a parameter region that is quite achievable in realistic experimental settings. This phase has demonstrated remarkable robustness against asymmetries, particularly large number imbalance. It points to new ways of realizing novel incompressible densities at fractional fillings with complementary fillings between the species of atoms as the result of strong anti-correlation. Our calculation also shows several possible measurements to detect the MI phase. These calculations are based on previous experiments on MI phases and we considered system sizes comparable to realistic experimental systems. In addition to the Fermi mixtures, similar MI phases can exist for mixtures of heavy-light Bose-Fermi and Bose-Bose atoms in the region where the intra-species bosonic interaction is stronger than the inter-species interaction.  
\begin{acknowledgements}
J.K.F. was supported by a MURI grant from the Air
Force Office of Scientific Research numbered FA9559-09-1-
0617 and by the McDevitt bequest at Georgetown University. M.M.M. acknowledges support by the Polish National Science Center (NCN) under grant DEC-2013/11/B/ST3/00824. This work was supported by the National Science Foundation under Physics Frontier Center grant PHY-0822671.
\end{acknowledgements}

\bibliography{FK_2}
\end{document}